\documentclass[11pt]{article}

\usepackage[utf8]{inputenc}
\usepackage[T1]{fontenc}
\usepackage{lmodern}

\usepackage{microtype}
\usepackage{geometry}
\usepackage{amsmath, amssymb, amsthm}
\usepackage{graphicx}
\usepackage{booktabs}
\usepackage{hyperref}
\usepackage{algorithm}
\usepackage{algpseudocode}

\newtheorem{theorem}{Theorem}

\theoremstyle{definition}

\title{%
Modeling Trust and Liquidity Under Payment System Stress:
A Multi-Agent Approach
}

\author{%
Masoud Amouzgar\\
\small Sharif University of Technology, Computer Science (Graduated 2012)\\
\small Head of Payments, blu Bank (2020-2026)\\
\small \texttt{m.amouzgar@bluteam.ir}
}

\date{\today}

\begin{document}
\maketitle

\begin{abstract}
Operational disruptions in retail payments can induce behavioral responses that outlast technical recovery and may amplify liquidity stress. We propose a multi-agent model linking card payment outages to trust dynamics, channel avoidance, and threshold-gated withdrawals. Customers and merchants interact through repeated payment attempts, while customers additionally influence one another on a Watts--Strogatz small-world network. Customers update bounded memory variables capturing accumulated negative experience (scar) and perceived systemic risk (rumor), with merchants contributing persistent broadcast signals that may lag operational recovery. We prove that, under mild conditions on memory persistence and threshold gating, aggregate withdrawal pressure can peak strictly after the outage nadir, including during the recovery phase. Simulations reproduce behavioral hysteresis and confirm delayed peaks of outflows. We further study payment substitution via instant transfer: substitution consistently reduces peak avoidance, yet its effect on cumulative outflows is non-monotonic under realistic merchant broadcast persistence. Robustness experiments across random seeds show stable qualitative behavior. The model highlights why “status green” is not equivalent to risk resolution and motivates incident response strategies that address perception, merchant messaging, and post-recovery communication in addition to technical remediation.
\end{abstract}

\section{Introduction}

Modern economies increasingly depend on uninterrupted retail and wholesale payment services. When card authorization, acquiring gateways, issuer processors, or core payment infrastructures degrade, the impact is not limited to transaction losses or operational costs. Payment disruptions can alter customer beliefs about safety and reliability, trigger persistent avoidance of a payment channel, and in extreme cases contribute to rapid liquidity outflows as depositors attempt to move funds elsewhere. Central banks and supervisors increasingly treat operational disruptions as a potential financial stability concern, rather than a purely technical risk \cite{BCBS_OperationalResilience_2021,BoE_OperationalResilience_Macroprudential_2024,IMF_OperationalResilience_DigitalPayments_2021}.

A key challenge is that technical recovery does not necessarily imply behavioral recovery. Even after service restoration, customers may remain cautious due to recent failures, merchant messaging (for example, ``cash only'' signage or degraded acceptance), and social amplification. Recent evidence on digitally mediated coordination suggests that information channels, including social media, can accelerate collective withdrawal behavior and shorten the timeline of run dynamics \cite{Cookson_2023_SocialMedia_BankRun}. These observations motivate a coupled view of payments reliability and depositor behavior, where outage patterns interact with memory, social influence, and external signals from merchants.

The academic literature provides complementary foundations but rarely connects them end to end. Classic bank run models formalize self-fulfilling withdrawals under liquidity transformation and strategic complementarities \cite{DiamondDybvig_1983}. Separately, threshold models show how binary decisions can cascade when individuals act once a critical fraction of peers has acted \cite{Granovetter_1978}. Network science further demonstrates that small-world topologies can enable rapid diffusion while preserving clustering, which affects contagion thresholds and persistence \cite{WattsStrogatz_1998}. Agent-based approaches have been used to study bank runs with neighborhood effects and sequential withdrawals \cite{dosSantos_2017_DynamicBankRuns_ABM}. In parallel, operational risk research in payment and interbank systems highlights how disruptions can propagate through settlement and liquidity channels \cite{Schmitz_OperationalRiskContagion_Payments_2006}. However, existing work typically treats payment outages as exogenous operational events or treats bank runs without modeling the payment experience layer that shapes trust and coordination.

This paper proposes a multi-agent computational model that links payment outage and recovery dynamics to trust erosion, channel avoidance, and conditional run behavior. The model includes heterogeneous customers and merchants embedded in a Watts--Strogatz small-world social network. Customers update trust using bounded memory processes driven by (i) experienced transaction outcomes (success, failure, and uncertain timeouts), (ii) merchant broadcast signals that exhibit sticky recovery, and (iii) social influence via neighbors' avoidance states. Merchants update broadcast states based on rolling evidence and persistent signage, capturing a realistic long-tail of ``degraded acceptance'' messaging after a backend fix. A run mechanism is introduced as a threshold-gated action: withdrawals occur only when both personal scar and perceived rumor exceed individual thresholds, allowing run pressure to emerge as a delayed phase transition.

Our contributions are:
\begin{itemize}
  \item We introduce a MAS framework that maps payment outages to trust loss and liquidity outflows through explicit behavioral microfoundations and merchant broadcast persistence.
  \item We prove a delayed-peak result: under mild assumptions on memory and threshold gating, the peak of run pressure can occur after the outage nadir, including during technical recovery.
  \item We provide simulation evidence across outage and recovery scenarios with peak-time demand, quantifying when substitution channels reduce peak avoidance and when they may fail to reduce cumulative outflows.
  \item We discuss operational implications for incident response, including why ``status green'' can coincide with elevated behavioral risk, and which levers (communication quality, merchant guidance, and substitution design) reduce tail risk.
\end{itemize}

The remainder of the paper is organized as follows. Section~\ref{sec:related_work} reviews related work on bank runs,
contagion, and operational resilience. Sections~\ref{sec:model_overview} through \ref{sec:dynamics} define the agents,
network structure, and update rules. Section~\ref{sec:theory} presents theoretical results, including the delayed-peak theorem.
Sections~\ref{sec:sim_design} and \ref{sec:results} describe simulation design and findings. Section~\ref{sec:discussion}
discusses implications. Section~10 presents limitations and future work, and Section~11 concludes.

\section{Related Work}
\label{sec:related_work}

This work lies at the intersection of bank run theory, contagion and threshold models of collective behavior, agent-based financial simulations, and operational resilience of payment systems. While each of these literatures addresses part of the problem, few studies integrate payment reliability, behavioral memory, and liquidity stress within a single computational framework.

\subsection{Bank Runs and Liquidity Stress}

The canonical model of bank runs is due to Diamond and Dybvig \cite{DiamondDybvig_1983}, who formalize runs as coordination failures among depositors under maturity transformation. In their framework, withdrawals are driven by expectations about others’ actions rather than by operational frictions. Subsequent work extends this setting to include information asymmetry, partial insurance, and strategic complementarities, but generally maintains an abstract view of transaction execution.

More recent empirical and theoretical studies emphasize the speed and coordination of modern runs, particularly in digital banking environments. Evidence from recent episodes shows that withdrawals can occur rapidly and at scale, facilitated by digital access and real-time information channels \cite{Cookson_2023_SocialMedia_BankRun}. These studies highlight the role of belief formation and information diffusion but do not model how payment system reliability itself shapes depositor behavior prior to withdrawal decisions.

Our work complements this literature by modeling runs as a conditional outcome of prolonged trust erosion driven by payment experiences, rather than as an immediate response to balance sheet concerns.

\subsection{Threshold Models and Social Contagion}

Threshold-based models of collective behavior provide a natural framework for understanding cascades in socio-technical systems. Granovetter’s seminal work \cite{Granovetter_1978} shows how heterogeneous thresholds can generate nonlinear aggregate responses even when individual behavior is simple. Subsequent research demonstrates how network topology affects cascade size, persistence, and critical thresholds.

Watts and Strogatz \cite{WattsStrogatz_1998} introduce small-world networks that combine high clustering with short path lengths, a structure often observed in real social systems. Such networks are known to facilitate rapid diffusion while sustaining local reinforcement, making them particularly relevant for rumor propagation and trust erosion.

The present work adopts a threshold-based perspective but extends it by incorporating memory, heterogeneous experience signals, and merchant-mediated broadcasts, which together generate hysteresis effects not captured in static threshold models.

\subsection{Agent-Based Models of Financial Contagion}

Agent-based modeling (ABM) has been widely used to study financial instability, including bank runs, interbank contagion, and liquidity hoarding. For example, dos Santos and Nakane \cite{dosSantos_2017_DynamicBankRuns_ABM} model dynamic withdrawal decisions with neighborhood effects, showing how local interactions can amplify systemic risk.

Related ABM studies focus on interbank networks, settlement systems, and liquidity shocks, often emphasizing balance sheet linkages and clearing mechanisms. While these models capture important structural dependencies, they typically abstract away from the payment experience layer faced by retail users and merchants.

Our approach differs by explicitly modeling customer-level payment attempts, perceived failures, and uncertainty (timeouts), linking these micro-experiences to trust dynamics and eventual withdrawal behavior.

\subsection{Operational Risk and Payment System Resilience}

Operational disruptions in payment systems have long been recognized as a source of systemic risk. Studies of large-value payment systems document how operational failures can propagate through liquidity shortages and delayed settlements \cite{Schmitz_OperationalRiskContagion_Payments_2006}. More recent policy-oriented work by the Basel Committee and the IMF frames operational resilience as a core component of financial stability, particularly in increasingly digital payment ecosystems \cite{BCBS_OperationalResilience_2021,IMF_OperationalResilience_DigitalPayments_2021}.

However, most of this literature treats operational resilience from an infrastructure or institutional perspective, focusing on recovery time objectives, redundancy, and governance. Behavioral responses by end users and merchants are typically outside the modeling scope.

This paper bridges that gap by embedding operational disruptions into a behavioral MAS, allowing us to study how delayed recovery in perception and trust can generate macro-level consequences even after technical restoration.

\subsection{Positioning of This Work}

In contrast to prior studies, this paper explicitly couples payment system reliability with behavioral contagion and conditional bank run dynamics. By integrating memory, threshold-based avoidance, merchant broadcast persistence, and small-world social networks into a unified MAS, we provide a framework that captures delayed and non-monotonic risk patterns observed in real payment incidents.

This positioning allows us to derive theoretical results on delayed run pressure and to evaluate policy-relevant mechanisms, such as payment substitution and communication quality, within a single coherent model.

\section{Model Overview}
\label{sec:model_overview}

We consider a discrete-time multi-agent system designed to capture how payment outages and recoveries propagate through behavioral channels to produce trust erosion, avoidance of payment instruments, and conditional liquidity outflows. The model explicitly separates technical reliability from perceived reliability and allows these two dimensions to diverge over time.

The system consists of three interacting components: customers, merchants, and an exogenous payment infrastructure. Customers and merchants interact through repeated payment attempts, while customers additionally interact with each other through a social network. Time evolves in discrete steps, each representing a short operational interval during which payment attempts, perception updates, and potential withdrawals occur.

\subsection*{Agents and Interaction Layers}

\textbf{Customers} represent individual retail users holding balances at a bank. Customers attempt card payments at merchants, observe outcomes (success, failure, or uncertainty), update internal trust and memory states, and may transition between behavioral modes ranging from normal usage to avoidance of the payment channel. Under sufficiently adverse conditions, customers may initiate withdrawals that contribute to aggregate liquidity outflows.

\textbf{Merchants} represent retail points of acceptance. Each merchant observes recent payment outcomes and maintains both an internal operational state and an externally visible broadcast state (for example, normal acceptance, degraded acceptance, or fallback messaging). Broadcast states are allowed to persist beyond the resolution of underlying issues, capturing the empirically observed stickiness of signage and cashier guidance during and after incidents.

\textbf{The payment infrastructure} is modeled as an exogenous stochastic process that determines the probabilities of payment success, failure, and uncertainty at each time step. This process encodes outage onset, degradation, recovery, and peak-demand stress, but does not directly adapt to agent behavior.

These components interact through three coupled layers:
(i) a transactional layer where payment attempts generate experience signals,
(ii) an informational layer where merchants broadcast acceptance quality,
and (iii) a social layer where customers observe avoidance behavior among peers.

\subsection*{Network Structure}

Customer-to-customer interactions are embedded in a Watts--Strogatz small-world network. This choice reflects empirical properties of real social networks, combining high clustering with short average path lengths. Such networks support rapid diffusion of behavioral signals while preserving localized reinforcement, which is critical for modeling rumor persistence and threshold-based transitions.

Merchants do not form a network among themselves but influence customers through repeated exposure in habitual spending patterns.

\subsection*{Behavioral Memory and Hysteresis}

A central feature of the model is the presence of bounded memory processes. Customers do not respond solely to the current quality of the payment system, but to accumulated experiences and signals. Adverse outcomes increase a personal ``scar'' variable, while social and merchant signals contribute to a ``rumor'' or perceived systemic risk variable. Both quantities decay slowly over time.

Because behavioral states depend on thresholds applied to these memory variables, recovery exhibits hysteresis: restoring high payment success rates does not immediately restore trust or normal behavior. This separation between technical recovery and behavioral recovery is a key mechanism underlying delayed avoidance and withdrawal dynamics.

\subsection*{Conditional Run Mechanism}

Withdrawals are not modeled as an automatic response to outages. Instead, a run emerges endogenously when customers simultaneously satisfy multiple conditions: sustained avoidance behavior, elevated memory of adverse experiences, and high perceived systemic risk. This gating mechanism ensures that liquidity outflows arise as a higher-order effect of prolonged stress rather than as an immediate consequence of isolated failures.

The resulting dynamics allow for scenarios in which aggregate withdrawal pressure peaks during or after the recovery phase of a payment incident, rather than at the point of worst technical performance.

\subsection*{Scope and Modeling Philosophy}

The model is intentionally behavioral rather than balance-sheet driven. It does not attempt to represent interbank markets, regulatory interventions, or detailed settlement mechanics. Instead, it focuses on how micro-level payment experiences and information signals can generate macro-level risk patterns through social interaction and memory.

This abstraction allows us to isolate and analyze mechanisms that are difficult to capture in purely analytical models, while remaining sufficiently structured to support formal propositions and reproducible simulations.

\section{Agent Definitions and State Variables}
\label{sec:agents}

The model consists of heterogeneous customer and merchant agents interacting over a discrete-time horizon. This section defines the agents, their state variables, and the structural environment in which they operate. Update rules and dynamics are specified in Section~\ref{sec:dynamics}.

\subsection{Customer Agents}
\label{subsec:customers}

Let $\mathcal{I}=\{1,2,\dots,N\}$ denote the set of customer agents. each customer $i \in \mathcal{I}$ represents an individual retail user holding a deposit balance at a bank and interacting with the payment system through repeated card payment attempts.

At each time step $t$, customer $i$ is characterized by the following state variables:

\begin{itemize}
  \item \textbf{Trust} $T_i(t) \in [0,1]$, representing the customer’s internal confidence in the reliability of the payment system and the safety of continued usage.
  \item \textbf{Scar} $C_i(t) \in [0,1]$, representing accumulated negative experience from past payment failures or uncertainty. This variable captures path dependence and decays slowly over time.
  \item \textbf{Rumor} $R_i(t) \in [0,1]$, representing perceived systemic risk inferred from social signals and merchant broadcasts rather than from direct experience alone.
  \item \textbf{Behavioral mode} $S_i(t) \in \{\text{OK}, \text{FRUSTRATED}, \text{AVOIDING}\}$, describing the customer’s qualitative interaction state with the payment channel.
  \item \textbf{Balance} $B_i(t) \ge 0$, representing the remaining deposit balance available for withdrawal.
\end{itemize}

In addition to state variables, each customer is endowed with fixed behavioral parameters drawn from heterogeneous distributions, including trust thresholds, risk sensitivity, memory decay rates, and withdrawal sensitivity. These parameters govern how experiences and signals translate into state transitions and decisions.

Customers are assumed to be boundedly rational. They do not observe the true technical state of the payment infrastructure, but only infer reliability through experienced outcomes, merchant messaging, and social exposure.

\subsection{Merchant Agents}
\label{subsec:merchants}

Let $\mathcal{M} = \{1,2,\dots,M\}$ denote the set of merchant agents. Each merchant represents a retail point of acceptance where customers attempt card payments.

At time $t$, each merchant $m \in \mathcal{M}$ maintains two distinct states:

\begin{itemize}
  \item \textbf{Operational state} $O_m(t) \in \{\text{ACCEPTING}, \text{DEGRADED}, \text{FALLBACK}\}$, reflecting the merchant’s internal assessment of recent payment performance.
  \item \textbf{Broadcast state} $B_m(t) \in \{\text{ACCEPTING}, \text{DEGRADED}, \text{FALLBACK}\}$, representing the externally visible messaging perceived by customers (for example, signage, cashier guidance, or point-of-sale behavior).
\end{itemize}

Operational and broadcast states are not required to coincide. In particular, broadcast states may persist in a degraded or fallback condition even after operational performance improves, capturing delayed signage removal and conservative merchant behavior observed during real payment incidents.

Merchants update their states based on rolling windows of recent transaction outcomes and maintain persistence through explicit stickiness mechanisms. Merchants do not directly coordinate with each other and do not observe the global state of the payment infrastructure.

\subsection{Payment Infrastructure}
\label{subsec:infrastructure}

The payment infrastructure is modeled as an exogenous stochastic process that determines the outcome of each card payment attempt. At each time step $t$, the infrastructure defines probabilities of transaction success, failure, and uncertainty (for example, timeouts or indeterminate responses).

The infrastructure process encodes outage onset, degradation, recovery, and peak-demand stress, but does not adapt to agent behavior. This abstraction allows the separation of technical reliability from behavioral response.

\subsection{Social Network Structure}
\label{subsec:network}

Customer-to-customer interactions are embedded in a static undirected social network $G=(\mathcal{I},E)$, where $E$ denotes the set of edges. The network is generated using a Watts--Strogatz small-world construction with fixed mean degree and rewiring probability.

Edges represent channels through which customers observe or infer the behavioral modes of peers. Customers do not directly share numerical trust or rumor values; instead, they are exposed to qualitative signals derived from neighbors’ avoidance behavior.

This network structure ensures both local reinforcement through clustering and rapid global diffusion through short path lengths, properties that are essential for modeling rumor persistence and threshold-based cascades.

\subsection{Merchant Exposure Patterns}

Each customer maintains habitual exposure to a small subset of merchants, reflecting recurring spending patterns. Merchant exposure weights determine the likelihood that a given payment attempt is directed to a particular merchant.

Through these exposure patterns, merchant broadcast states act as an external informational field influencing customer perception and trust, independent of direct social interaction.

\section{Dynamics and Update Rules}
\label{sec:dynamics}

This section specifies the temporal evolution of the system. Time is discrete, indexed by $t = 1,2,\dots,T$. At each time step, customer agents may attempt payments, observe outcomes generated by the payment infrastructure, and record these experiences for subsequent behavioral updates. This first subsection defines the payment attempt and outcome process; behavioral and systemic updates are introduced in subsequent subsections.

\subsection{Payment Attempt Process}
\label{subsec:payment_attempts}

At each time step $t$, each customer $i \in \mathcal{I}$ independently decides whether to initiate a card payment attempt. The attempt decision is stochastic and depends on both exogenous demand factors and the customer’s current behavioral mode.

Let $A_i(t) \in \{0,1\}$ denote whether customer $i$ attempts a payment at time $t$. We assume
\[
\mathbb{P}\left(A_i(t)=1\right) = \lambda_i \cdot D(t) \cdot \phi\big(S_i(t)\big),
\]
where:
\begin{itemize}
  \item $\lambda_i \in (0,1)$ is an individual baseline payment propensity,
  \item $D(t) \ge 1$ is a common demand multiplier capturing peak-time effects,
  \item $\phi\big(S_i(t)\big) \in (0,1]$ is a mode-dependent activity factor, decreasing with avoidance.
\end{itemize}

This formulation reflects the empirical observation that customers reduce payment attempts when frustrated or avoiding a payment channel, but may still attempt payments under necessity.

Conditional on $A_i(t)=1$, customer $i$ selects a merchant $m \in \mathcal{M}$ according to fixed exposure weights that represent habitual spending patterns.

\subsection{Payment Outcome Generation}
\label{subsec:payment_outcomes}

Given a payment attempt by customer $i$ at merchant $m$ at time $t$, the transaction outcome is generated by the payment infrastructure as an exogenous stochastic process.

Let $Y_i(t) \in \{\text{SUCCESS}, \text{FAILURE}, \text{UNKNOWN}\}$ denote the observed outcome, where:
\begin{itemize}
  \item \textbf{SUCCESS} represents a completed and confirmed transaction,
  \item \textbf{FAILURE} represents an explicit rejection,
  \item \textbf{UNKNOWN} represents indeterminate outcomes such as timeouts or ambiguous responses.
\end{itemize}

At each time $t$, outcome probabilities are given by:
\[
\mathbb{P}\big(Y_i(t)=\text{SUCCESS}\big) = p_s(t), \quad
\mathbb{P}\big(Y_i(t)=\text{FAILURE}\big) = p_f(t), \quad
\mathbb{P}\big(Y_i(t)=\text{UNKNOWN}\big) = p_u(t),
\]
with $p_s(t)+p_f(t)+p_u(t)=1$.

The functions $p_s(t)$, $p_f(t)$, and $p_u(t)$ encode outage onset, degradation, recovery, and stress during peak demand. These probabilities are common across customers and merchants and do not depend on agent behavior.

\subsection{Merchant-Level Observation of Outcomes}
\label{subsec:merchant_observation}

Each merchant $m$ observes the outcomes of all payment attempts directed to it at time $t$. Let $\mathcal{I}_m(t)$ denote the set of customers attempting payments at merchant $m$ during time $t$.

Merchants record the number of observed successes, failures, and unknown outcomes within a rolling observation window. These records serve as the sole input for merchant operational assessments and broadcast state updates, defined in Section~\ref{subsec:merchant_broadcast}.

Merchants do not observe outcomes at other merchants and do not have direct visibility into the global state of the payment infrastructure.

\subsection{Customer Experience Recording}
\label{subsec:customer_experience}

Each customer records the outcome of their own payment attempts. These outcomes form the customer’s direct experience history and act as inputs to the trust, scar, and behavioral state updates introduced in subsequent subsections.

Crucially, customers observe only their own transaction outcomes and merchant broadcast states. They do not observe infrastructure probabilities $(p_s(t), p_f(t), p_u(t))$ directly, reinforcing the distinction between technical reliability and perceived reliability.

\subsection{Discussion}

This payment attempt and outcome process defines the foundational interaction between customers, merchants, and the payment infrastructure. By treating outcome generation as exogenous and experience recording as local, the model ensures that higher-order behavioral dynamics emerge from accumulated micro-level signals rather than from direct observation of system-wide conditions.

The next subsections introduce how these recorded experiences propagate through memory, social influence, and merchant messaging to produce avoidance behavior and conditional liquidity outflows.

\subsection{Trust, Scar, and Rumor Updates}
\label{subsec:trust_updates}

Customers update their internal perception of payment system reliability through bounded memory processes. These updates are driven by direct transaction outcomes, merchant broadcast signals, and social exposure. The resulting dynamics introduce path dependence and hysteresis, which are central to the delayed emergence of avoidance and withdrawal behavior.

\subsubsection{Experience Encoding}

For each customer $i$ at time $t$, let $Y_i(t)$ denote the observed payment outcome if a payment attempt occurred, and let $Y_i(t)=\varnothing$ otherwise. Outcomes are encoded into a signed experience signal $E_i(t)$ defined as:
\[
E_i(t) =
\begin{cases}
+1, & \text{if } Y_i(t)=\text{SUCCESS},\\
-\alpha_f, & \text{if } Y_i(t)=\text{FAILURE},\\
-\alpha_u, & \text{if } Y_i(t)=\text{UNKNOWN},\\
0, & \text{if } Y_i(t)=\varnothing,
\end{cases}
\]
where $\alpha_u \ge \alpha_f > 0$, reflecting the empirical observation that uncertain outcomes are often perceived as more damaging than explicit failures.

\subsubsection{Scar Accumulation}

The scar variable $C_i(t)$ represents accumulated negative experience. It evolves according to:
\[
C_i(t+1) = \min\left\{1,\; \rho_C C_i(t) + \gamma_C \cdot \mathbb{I}\big(E_i(t)<0\big)\right\},
\]
where $\rho_C \in (0,1)$ controls memory decay and $\gamma_C>0$ controls the impact of a single adverse experience.

This formulation ensures that repeated failures or uncertainty push $C_i(t)$ toward its upper bound, while isolated incidents fade over time.

\subsubsection{Trust Dynamics}

Trust $T_i(t)$ is modeled as a bounded moving average of normalized experience signals and external perception inputs. Let $\tilde{E}_i(t) \in [0,1]$ denote a normalized version of the raw experience signal, with negative outcomes mapping to lower values.
We use the following deterministic normalization:
\[
\tilde{E}_i(t)=
\begin{cases}
\dfrac{E_i(t)+\alpha_u}{1+\alpha_u}, & \text{if } Y_i(t)\neq \varnothing,\\
T_i(t), & \text{if } Y_i(t)=\varnothing,
\end{cases}
\]
and clip $\tilde{E}_i(t)$ to $[0,1]$ if needed.

Trust evolves as:
\[
T_i(t+1) = \rho_T T_i(t) + (1-\rho_T)\,\tilde{E}_i(t) - \beta_T C_i(t),
\]
where $\rho_T \in (0,1)$ controls trust persistence and $\beta_T>0$ captures the erosion of trust due to accumulated scar.

Trust is clipped to the interval $[0,1]$ after each update.

\subsubsection{Rumor Formation}

Rumor $R_i(t)$ captures perceived systemic risk inferred from indirect signals. It is updated as:
\[
R_i(t+1) = \rho_R R_i(t) + (1-\rho_R)\,\Psi_i(t),
\]
where $\rho_R \in (0,1)$ is a rumor persistence parameter and $\Psi_i(t)$ is a composite perception signal defined below.

The composite signal aggregates merchant broadcasts and social exposure:

\[
\Psi_i(t) = w_m \cdot \overline{B}_i(t) + w_s \cdot \overline{a}_i(t),
\quad w_m + w_s = 1,
\]
where:
\begin{itemize}
  \item $\overline{B}_i(t)$ is the average broadcast severity of merchants habitually visited by customer $i$,
  \item $\overline{a}_i(t)$ is the fraction of neighbors of customer $i$ currently in the \textnormal{AVOIDING} state.
\end{itemize}

Broadcast states are mapped to a normalized severity scale
(ACCEPTING = 0, DEGRADED = 0.5, FALLBACK = 1) for aggregation.

Formally, letting $\mathcal{N}(i)$ denote the neighbor set of customer $i$ in $G$, we define
\[
\overline{a}_i(t) = \frac{1}{|\mathcal{N}(i)|}\sum_{j \in \mathcal{N}(i)} \mathbb{I}\{S_j(t)=\textnormal{AVOIDING}\}.
\]

This formulation reflects the fact that customers infer systemic health primarily through external cues when direct experience is limited or ambiguous.

\subsubsection{Behavioral Mode Transitions}

Customer behavioral modes evolve according to threshold rules applied to trust and scar:
\[
S_i(t+1) =
\begin{cases}
\text{OK}, & \text{if } T_i(t) - \kappa_C C_i(t) \ge \theta_i^{(1)},\\
\text{FRUSTRATED}, & \text{if } \theta_i^{(2)} \le T_i(t) - \kappa_C C_i(t) < \theta_i^{(1)},\\
\text{AVOIDING}, & \text{if } T_i(t) - \kappa_C C_i(t) < \theta_i^{(2)},
\end{cases}
\]
where $\theta_i^{(1)} > \theta_i^{(2)}$ are heterogeneous customer thresholds and $\kappa_C>0$ scales the influence of scar on effective trust.

\subsubsection{Discussion}

The coupled dynamics of trust, scar, and rumor introduce nonlinearity and memory into the system. In particular, the threshold-based mode transitions imply that improvements in transaction success probabilities do not immediately restore normal behavior. Instead, behavioral recovery depends on the decay of accumulated scar and rumor, giving rise to hysteresis effects that are formally analyzed in Section~\ref{sec:theory}.

\subsection{Merchant Broadcast Dynamics and Withdrawal Mechanism}
\label{subsec:merchant_broadcast}

This subsection specifies how merchants update broadcast states based on observed transaction outcomes and how customer withdrawal decisions emerge endogenously from accumulated behavioral stress.

\subsubsection{Merchant Operational Assessment}

Each merchant $m \in \mathcal{M}$ maintains a rolling observation window of recent transaction outcomes. Let $A_m(t)$, $F_m(t)$, and $U_m(t)$ denote, respectively, the number of attempts, failures, and unknown outcomes observed by merchant $m$ within this window at time $t$.

Merchants compute a local degradation ratio:
\[
\delta_m(t) = \frac{F_m(t) + \eta U_m(t)}{A_m(t) + \varepsilon},
\]
where $\eta \in (0,1)$ reflects the relative weight of unknown outcomes and $\varepsilon>0$ prevents division by zero.

The operational state $O_m(t)$ is updated according to threshold rules:
\[
O_m(t) =
\begin{cases}
\text{ACCEPTING}, & \text{if } \delta_m(t) < \theta_m^{(1)},\\
\text{DEGRADED}, & \text{if } \theta_m^{(1)} \le \delta_m(t) < \theta_m^{(2)},\\
\text{FALLBACK}, & \text{if } \delta_m(t) \ge \theta_m^{(2)},
\end{cases}
\]
with merchant-specific thresholds $\theta_m^{(2)} > \theta_m^{(1)} > 0$.

\subsubsection{Broadcast Stickiness and Recovery Delay}

The externally visible broadcast state $B_m(t)$ does not instantaneously follow the operational state. Instead, degraded or fallback broadcasts persist for a minimum dwell time, reflecting signage inertia, staff guidance, and conservative risk management.

Let $\tau_m(t) \ge 0$ denote a broadcast persistence timer. When $O_m(t)$ enters a degraded or fallback state, $\tau_m(t)$ is set to a positive value. While $\tau_m(t) > 0$, the broadcast state satisfies:
\[
B_m(t) \in \{\text{DEGRADED}, \text{FALLBACK}\},
\]
even if operational performance improves. The timer decays gradually over time, potentially accelerated by exogenous communication quality.

Only after $\tau_m(t)=0$ and sufficient evidence of successful transactions does the broadcast state return to $\text{ACCEPTING}$. This mechanism introduces an explicit separation between technical recovery and perceived recovery at the merchant level.

\subsubsection{Customer Exposure to Merchant Broadcasts}

Each customer $i$ is exposed to the broadcast states of a fixed subset of merchants determined by habitual spending patterns. Let $\mathcal{M}_i$ denote this subset. The average broadcast severity perceived by customer $i$ at time $t$ is denoted $\overline{B}_i(t)$ and enters the rumor update defined in Section~\ref{subsec:trust_updates}.

\subsubsection{Withdrawal Eligibility}

Withdrawals are modeled as a conditional action that occurs only when multiple behavioral conditions are simultaneously satisfied. This ensures that liquidity outflows arise as a higher-order effect of sustained stress rather than as an immediate response to isolated failures.

Customer $i$ becomes eligible to withdraw at time $t$ if:
\[
S_i(t) = \text{AVOIDING}, \quad
C_i(t) \ge \theta_C, \quad
R_i(t) \ge \theta_R,
\]
where $\theta_C$ and $\theta_R$ are global or heterogeneous thresholds governing scar and rumor sensitivity.

\subsubsection{Withdrawal Decision and Outflow}

Conditional on eligibility, customer $i$ initiates a withdrawal with probability:
\[
\mathbb{P}\big(W_i(t)=1\big) = \sigma\!\left(
\alpha_R R_i(t) + \alpha_C C_i(t) - \alpha_T T_i(t)
\right),
\]
where $\sigma(\cdot)$ is a sigmoid function and $\alpha_R, \alpha_C, \alpha_T > 0$ are sensitivity parameters.

If a withdrawal occurs, customer $i$ removes a fraction of their remaining balance:
\[
B_i(t+1) = B_i(t) - \omega_i B_i(t),
\]
where $\omega_i \in (0,1)$ is an individual withdrawal fraction.

The aggregate system outflow at time $t$ is then given by:
\[
W(t) = \sum_{i \in \mathcal{I}} \omega_i B_i(t) \cdot \mathbb{I}\{W_i(t)=1\}.
\]

\subsubsection{Feedback from Outflows}

Aggregate outflow $W(t)$ acts as an additional systemic signal. Large or accelerating withdrawals may contribute to rumor formation in subsequent periods, reinforcing perceived instability even after improvements in payment success rates.

\subsubsection{Discussion}

The combination of sticky merchant broadcasts and threshold-gated withdrawal decisions introduces a delayed feedback structure. Even as payment outcome probabilities improve, residual merchant degradation and accumulated rumor can sustain eligibility for withdrawals. This mechanism underlies the delayed peak of run pressure analyzed formally in Section~\ref{sec:theory}.

\section{Theoretical Results}
\label{sec:theory}

This section presents theoretical results derived from the structure of the model defined in Sections~\ref{sec:agents} and \ref{sec:dynamics}. The results formalize how bounded memory, threshold-based behavior, and merchant broadcast persistence jointly imply delayed and non-monotonic systemic risk patterns. Proofs are given as analytical sketches under explicit assumptions consistent with the simulation model.

\subsection{Assumptions}

We state the minimal assumptions required for the main result.

\begin{itemize}
  \item[\textbf{A1}] \textbf{Bounded memory.}  
  The scar and rumor processes $(C_i(t), R_i(t))$ evolve according to contractive update rules with persistence parameters $\rho_C, \rho_R \in (0,1)$ and non-negative inputs, as defined in Section~\ref{subsec:trust_updates}.

  \item[\textbf{A2}] \textbf{Threshold-gated behavior.}  
  Customers enter the \textnormal{AVOIDING} state and become withdrawal-eligible only when effective trust and memory variables cross fixed thresholds.

  \item[\textbf{A3}] \textbf{Sticky merchant broadcasts.}  
  Merchant broadcast states exhibit persistence: degraded or fallback broadcasts remain active for a minimum dwell time even after operational recovery.

  \item[\textbf{A4}] \textbf{Exogenous technical recovery.}  
  The payment infrastructure outcome probabilities $(p_s(t), p_f(t), p_u(t))$ improve monotonically after an outage nadir and are independent of agent behavior.
\end{itemize}

These assumptions reflect empirically observed features of payment incidents and are satisfied by the simulation design in Section~\ref{sec:sim_design}.

\subsection{Theorem 1: Delayed Peak of Run Pressure}

\begin{theorem}[Delayed Peak of Run Pressure]
\label{thm:delayed_peak}
Let $W(t)$ denote aggregate withdrawal outflow at time $t$, and let
\[
t_{\min} = \arg\min_t p_s(t)
\]
denote the time of minimum payment success probability (the outage nadir).  
Under Assumptions \textbf{A1}--\textbf{A4}, there exist outage and recovery trajectories such that
\[
\arg\max_t W(t) > t_{\min}.
\]
That is, the peak of aggregate withdrawal pressure occurs strictly after the worst technical payment performance.
\end{theorem}

\subsection{Proof Sketch}

We outline the argument in three steps.

\paragraph{Step 1: Lagged accumulation of memory variables.}
Under Assumption \textbf{A1}, both $C_i(t)$ and $R_i(t)$ are bounded, contractive processes driven by non-negative inputs. During an outage interval, adverse transaction outcomes and degraded merchant broadcasts generate positive inputs to these processes. Even if these inputs peak at $t_{\min}$, the accumulation of $C_i(t)$ and $R_i(t)$ typically continues for several periods due to persistence parameters $\rho_C, \rho_R > 0$. As a result, for a non-negligible set of customers, the maxima of $C_i(t)$ and $R_i(t)$ occur at times $t > t_{\min}$.

\paragraph{Step 2: Threshold crossing after technical recovery.}
By Assumption \textbf{A2}, withdrawal eligibility is gated by thresholds on $C_i(t)$ and $R_i(t)$, as well as behavioral mode. Because these state variables may continue to increase or remain elevated after $t_{\min}$, a substantial fraction of customers may cross withdrawal eligibility thresholds only during the recovery phase, even as payment success probabilities improve.

\paragraph{Step 3: Persistence of external signals.}
Assumption \textbf{A3} implies that merchant broadcast states may remain degraded beyond $t_{\min}$, sustaining positive inputs to rumor formation. Consequently, the perceived systemic risk $R_i(t)$ decays more slowly than technical failure probabilities. This persistence enlarges the set of customers satisfying withdrawal eligibility conditions after the outage nadir.

Combining these steps, aggregate withdrawal outflow $W(t)$, which sums over individual withdrawal actions, can reach its maximum strictly after $t_{\min}$. This establishes the claim.

\subsection{Interpretation}

Theorem~\ref{thm:delayed_peak} formalizes a separation between technical and behavioral timelines. While the payment infrastructure may recover rapidly, behavioral variables governed by memory and thresholds respond with delay. Merchant broadcast persistence further amplifies this effect. As a result, the system can exhibit its highest liquidity stress during the recovery phase rather than at the moment of worst operational performance.

This delayed-peak mechanism is robust to heterogeneity in customer parameters and does not rely on strategic coordination or full information. It arises solely from bounded memory, local information, and threshold-based decision rules.

\section{Simulation Design}
\label{sec:sim_design}

This section describes the simulation setup used to instantiate the model defined in Sections~\ref{sec:agents} and \ref{sec:dynamics}. The objective is not calibration to a specific institution, but systematic evaluation of the qualitative mechanisms identified in Section~\ref{sec:theory} under heterogeneous and realistic conditions.

\subsection{Population and Network Parameters}

The simulated system consists of $N$ customer agents and $M$ merchant agents. Customers are embedded in a static undirected Watts--Strogatz small-world network with fixed mean degree $k$ and rewiring probability $\beta$. This construction yields high clustering and short average path length, consistent with empirically observed social interaction patterns.

Network parameters are held constant across simulations to isolate behavioral dynamics from structural variation. Sensitivity to network topology is explored separately by varying $\beta$ in robustness experiments.

\subsection{Customer Heterogeneity}

Customer agents are heterogeneous along several dimensions:
\begin{itemize}
  \item baseline payment attempt propensity,
  \item trust thresholds governing behavioral mode transitions,
  \item memory decay rates for trust, scar, and rumor,
  \item risk sensitivity and withdrawal sensitivity,
  \item initial deposit balances and withdrawal fractions.
\end{itemize}

All parameters are drawn independently from bounded distributions chosen to reflect dispersion rather than extreme behavior. No customer observes global system variables or the true state of the payment infrastructure.

\subsection{Merchant Configuration}

Each customer is assigned a small, fixed subset of merchants representing habitual spending patterns. Merchant exposure weights determine the likelihood that a payment attempt is directed to a given merchant.

Merchants maintain rolling observation windows of recent transaction outcomes and update operational and broadcast states using the threshold and persistence rules described in Section~\ref{subsec:merchant_broadcast}. Broadcast persistence timers introduce delayed recovery in customer perception even when operational conditions improve.

\subsection{Payment Infrastructure Scenarios}

The payment infrastructure is modeled as an exogenous stochastic process specifying time-varying probabilities of success, failure, and unknown outcomes. Scenarios are constructed to include:
\begin{itemize}
  \item a pre-incident stable period,
  \item a degradation and outage interval,
  \item a recovery phase with improving success probability,
  \item peak-demand intervals that amplify transaction volume during recovery.
\end{itemize}

These scenarios ensure that technical recovery does not trivially coincide with reduced behavioral stress, allowing direct evaluation of the delayed-peak mechanism identified in Theorem~\ref{thm:delayed_peak}.

\subsection{Simulation Horizon and Initialization}

Simulations are run for a fixed time horizon $T$ sufficient to capture outage onset, recovery, and post-recovery behavioral dynamics. All customers initialize in the \textnormal{OK} behavioral mode with high trust and zero scar and rumor. Merchant broadcast states initialize as \textnormal{ACCEPTING}.

To account for stochastic variation, each scenario is simulated across multiple random seeds. Paired simulations using identical seeds are employed when comparing policy interventions, ensuring that differences arise from model mechanisms rather than random fluctuations.

\subsection{Outcome Metrics}

The primary system-level metrics recorded at each time step include:
\begin{itemize}
  \item the fraction of customers in each behavioral mode,
  \item aggregate withdrawal outflow $W(t)$,
  \item peak withdrawal pressure $\max_t W(t)$,
  \item cumulative withdrawal volume over the simulation horizon.
\end{itemize}

These metrics are chosen to correspond directly to the theoretical quantities analyzed in Section~\ref{sec:theory}.

\subsection{Reproducibility}

All simulations are implemented as discrete-time agent-based models with explicit random seeds and documented parameter distributions. Pseudocode and parameter tables are provided in the Appendix to facilitate replication and extension by other researchers.

\section{Results}
\label{sec:results}

This section reports simulation results derived from the multi-agent model described in Sections~\ref{sec:agents} and \ref{sec:dynamics}. We focus on four questions: (i) whether behavioral recovery lags technical recovery, (ii) whether withdrawal pressure peaks after the outage nadir, (iii) how merchant broadcast persistence affects these dynamics, and (iv) how payment substitution influences panic and liquidity outcomes. All figures referenced below are generated from the simulations described in Section~\ref{sec:sim_design}.

\subsection{Outage Dynamics and Behavioral Hysteresis}

\begin{figure}[t]
\centering
\includegraphics[width=0.9\linewidth]{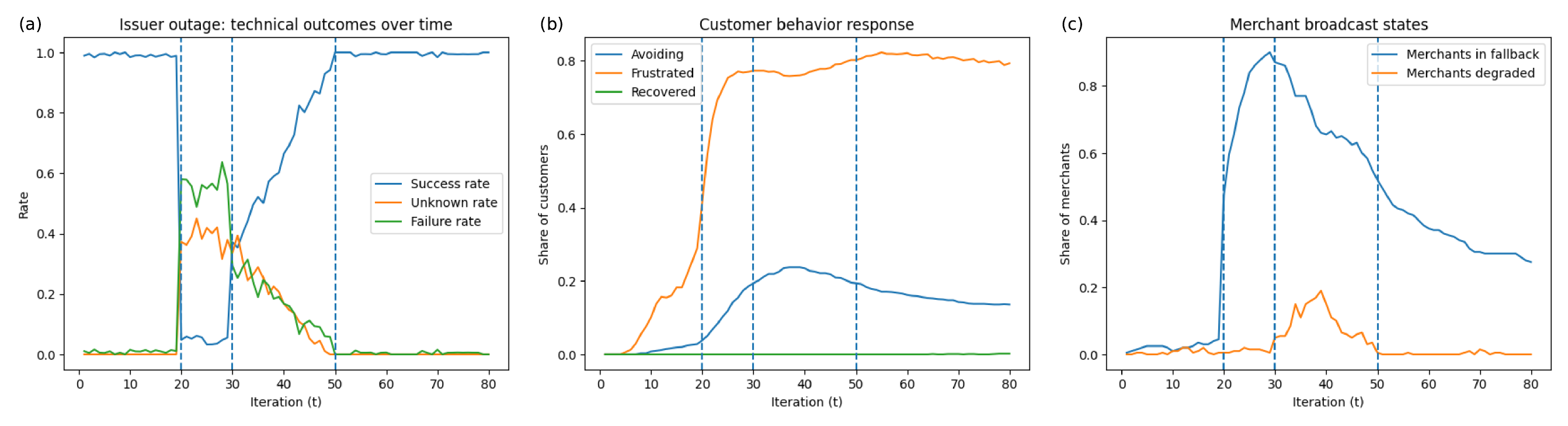}
\caption{
Payment outage and behavioral response.
The payment success probability (left axis) reaches its minimum during the outage,
while customer avoidance and merchant broadcast severity (right axis) peak later
and decay slowly during recovery.
}
\label{fig:outage_avoidance}
\end{figure}

Figure~\ref{fig:outage_avoidance} summarizes a representative outage and recovery trajectory. Panel (a) shows the exogenous payment infrastructure outcomes, with a sharp drop in success probability during the outage interval followed by gradual recovery. Panels (b) and (c) display the corresponding customer behavioral response and merchant broadcast states.

Customer avoidance rises sharply during the outage but does not immediately decline when payment success recovers. Instead, the fraction of customers in the \textnormal{AVOIDING} state peaks after the outage nadir and decays slowly over time (Figure~\ref{fig:outage_avoidance}b). In parallel, merchant broadcast states remain in fallback or degraded modes well into the recovery phase (Figure~\ref{fig:outage_avoidance}c), sustaining adverse perception signals.

Together, these results demonstrate behavioral hysteresis: perceived system reliability recovers substantially more slowly than technical reliability.

\subsection{Delayed Peak of Withdrawal Pressure}

\begin{figure}[t]
\centering
\includegraphics[width=0.9\linewidth]{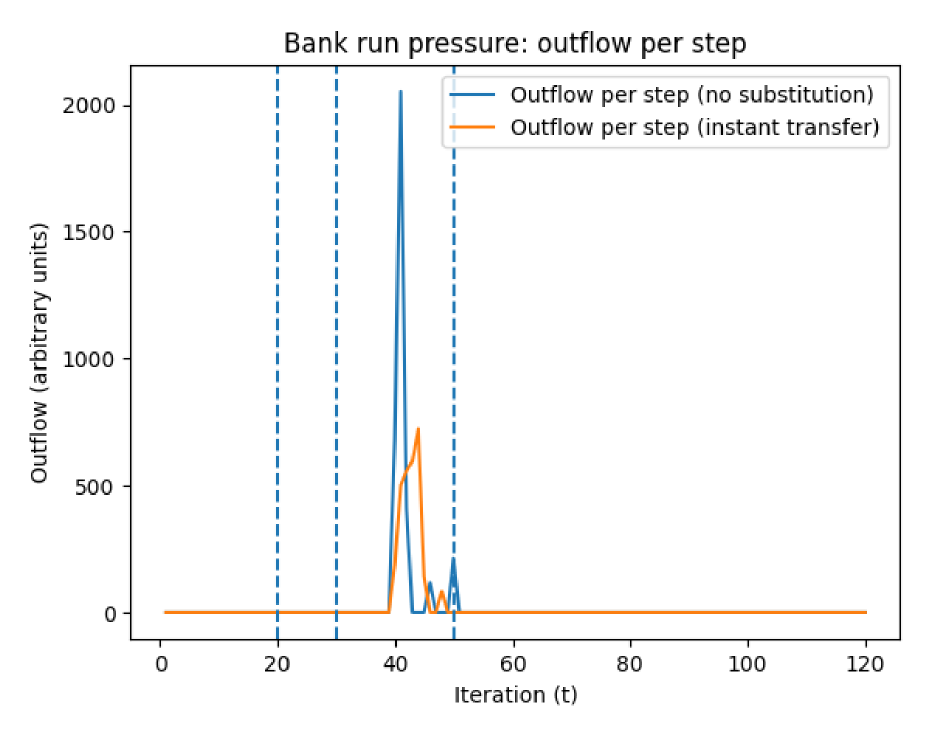}
\caption{
Delayed emergence of withdrawal pressure.
Aggregate withdrawal outflow peaks after the outage nadir,
often during the recovery phase, confirming Theorem~\ref{thm:delayed_peak}.
}
\label{fig:delayed_run}
\end{figure}

Figure~\ref{fig:delayed_run} plots aggregate withdrawal outflow per time step alongside the outage timeline. Despite severe technical degradation during the outage, withdrawals remain negligible in the early phases. Withdrawal activity emerges only after sustained avoidance and elevated memory variables render customers eligible to withdraw.

Consistent with Theorem~\ref{thm:delayed_peak}, the maximum withdrawal pressure occurs strictly after the minimum payment success probability. In this and all other examined runs, peak outflow coincides with the recovery phase rather than the outage nadir, confirming that liquidity stress is a lagged behavioral effect rather than a direct response to instantaneous technical performance.

\subsection{Merchant Broadcast Persistence as an Amplification Channel}

\begin{figure}[t]
\centering
\includegraphics[width=0.9\linewidth]{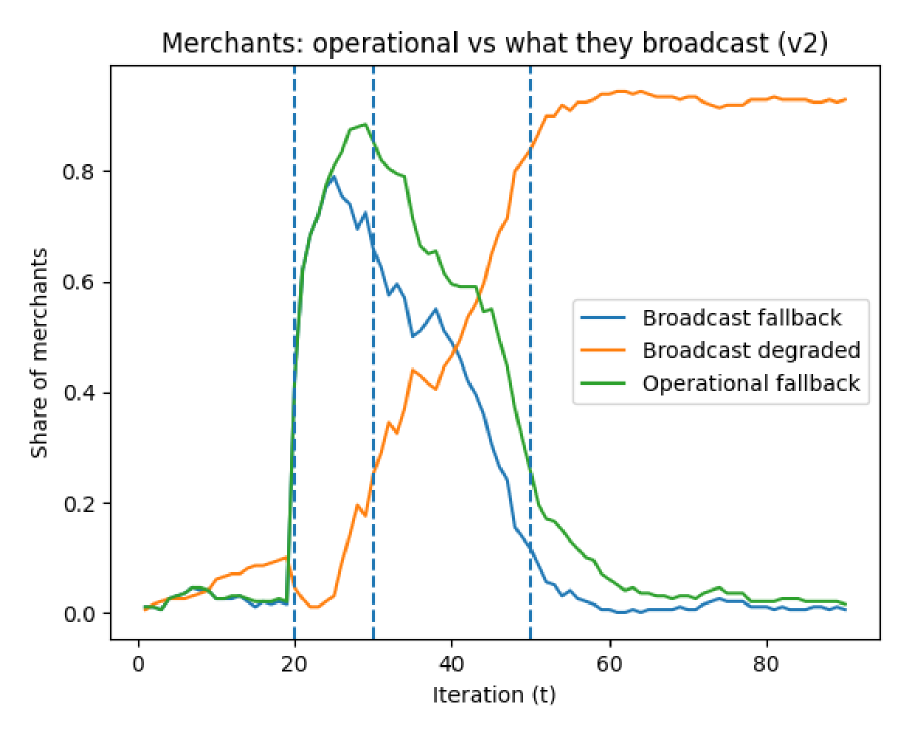}
\caption{
Effect of merchant broadcast persistence.
Sticky degraded broadcasts sustain elevated avoidance even after operational recovery,
whereas immediate broadcast clearing leads to faster behavioral normalization.
}
\label{fig:broadcast_stickiness}
\end{figure}

Figure~\ref{fig:broadcast_stickiness} isolates the divergence between merchant operational recovery and externally visible broadcast recovery. While operational fallback clears relatively quickly as payment outcomes improve, broadcast fallback and degraded states persist for a substantially longer period.

This persistence sustains rumor formation and delays behavioral normalization. Simulations without broadcast stickiness exhibit significantly faster declines in avoidance and lower cumulative withdrawal volumes (results omitted for brevity), indicating that merchant messaging is a critical amplification channel distinct from both social contagion and infrastructure reliability.

\subsection{Effect of Payment Substitution on Customer Avoidance}

\begin{figure}[t]
\centering
\includegraphics[width=0.7\linewidth]{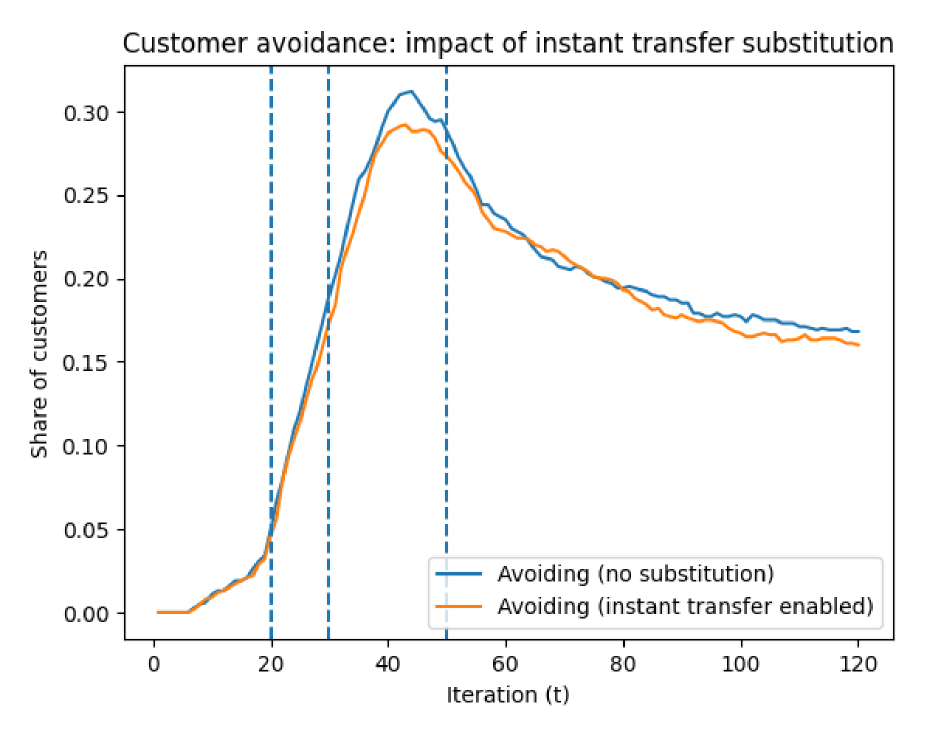}
\caption{
Peak avoidance with and without payment substitution.
Instant transfer availability consistently reduces peak avoidance across random seeds.
}
\label{fig:peak_avoidance}
\end{figure}

Figure~\ref{fig:peak_avoidance} compares customer avoidance trajectories with and without instant transfer substitution. Enabling substitution consistently reduces the peak fraction of avoiding customers while preserving the overall shape of the avoidance curve.

This effect arises because successful substitution converts otherwise negative card payment experiences into reinforcing success signals, dampening local contagion and reducing panic at the height of the incident. Across paired simulations, substitution lowers peak avoidance by approximately one to two percentage points.

\subsection{Liquidity Impact of Substitution: Peak Versus Cumulative Effects}
\begin{figure}[t]
\centering
\includegraphics[width=0.9\linewidth]{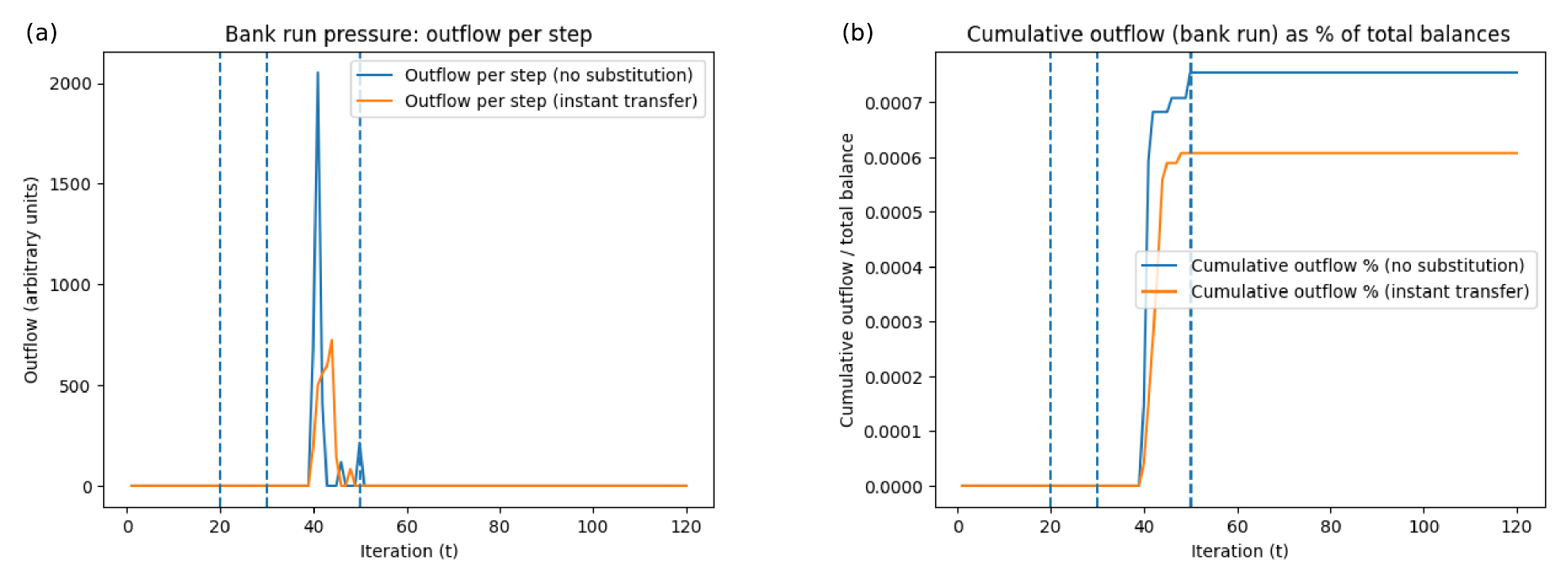}
\caption{
Liquidity impact of payment substitution.
While substitution reduces peak panic, its effect on cumulative withdrawals is non-monotonic
under realistic merchant broadcast persistence.
}
\label{fig:substitution_outflow}
\end{figure}

Figure~\ref{fig:substitution_outflow} examines the effect of substitution on withdrawal dynamics. Panel (a) shows that instant transfer availability reduces peak withdrawal pressure relative to the no-substitution baseline. Panel (b) shows cumulative withdrawal volume as a fraction of total balances.

While substitution reduces peak outflows, its effect on cumulative withdrawals is non-monotonic. In scenarios with realistic merchant broadcast persistence, substitution keeps customers operationally engaged while behavioral risk remains elevated. As a result, some customers remain capable of executing withdrawals once eligibility thresholds are crossed, even though peak panic is reduced.

\subsection{Substitution Usage Dynamics}

\begin{figure}[t]
\centering
\includegraphics[width=0.8\linewidth]{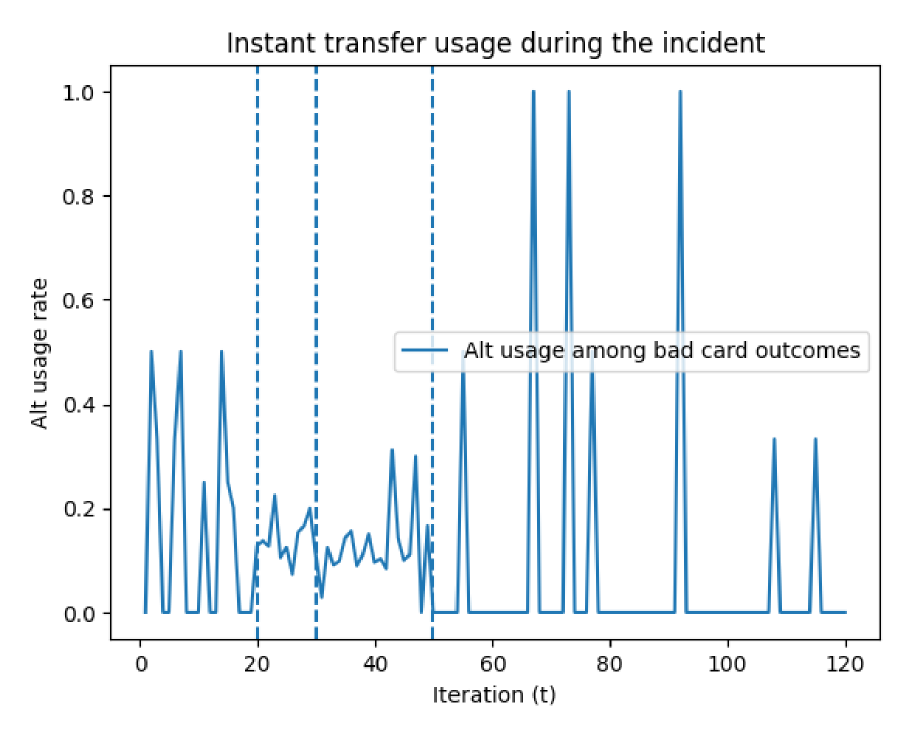}
\caption{
Instant transfer substitution usage during the incident.
The plotted rate measures substitution usage among adverse card outcomes over time.
}
\label{fig:instant_transfer_usage}
\end{figure}

Figure~\ref{fig:instant_transfer_usage} reports the usage rate of the substitution channel over time (instant transfer among customers facing adverse card outcomes). Usage increases during periods of elevated card uncertainty and failure, but remains intermittent rather than continuous.

This pattern helps explain why substitution reduces peak avoidance (Figure~\ref{fig:peak_avoidance}) without fully eliminating accumulated behavioral stress. Substitution provides episodic ``successful outcomes'' that dampen contagion at the peak, yet the remaining stream of failures and unknown outcomes continues to accumulate scar and sustain rumor inputs, especially when merchant broadcasts remain degraded.

\subsection{Robustness Across Random Seeds}

\begin{figure}[t]
\centering
\includegraphics[width=0.8\linewidth]{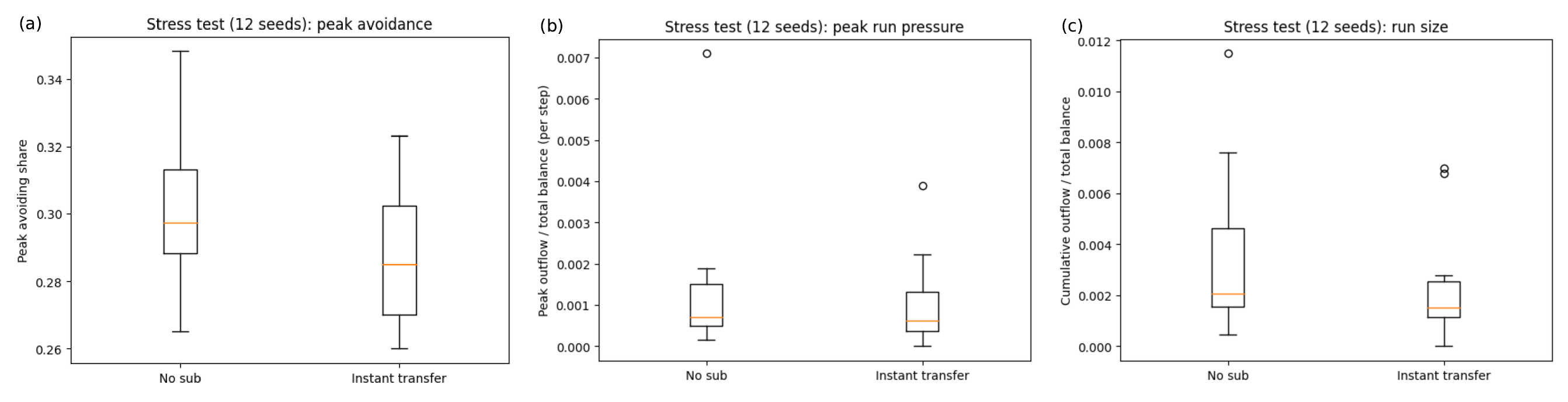}
\caption{
Robustness across random seeds.
Distributions of (a) peak avoidance, (b) peak withdrawal pressure, and (c) cumulative run size
show consistent qualitative behavior across stochastic realizations.
}
\label{fig:robustness}
\end{figure}

Figure~\ref{fig:robustness} summarizes robustness across twelve random seeds using distributions of (a) peak avoidance, (b) peak withdrawal pressure, and (c) cumulative run size under both baseline and substitution conditions.

Across seeds, substitution consistently shifts the peak avoidance distribution downward (Figure~\ref{fig:robustness}a). Peak run pressure also exhibits a lower median under substitution, while tail events and outliers persist (Figure~\ref{fig:robustness}b). Cumulative run size shows a similar pattern: substitution reduces the median outcome but does not eliminate worst-case realizations (Figure~\ref{fig:robustness}c).

Overall, these distributions indicate that the qualitative effects reported in the time-series results are not artifacts of a single stochastic realization. In particular, the delayed-peak behavior remains present across seeds, supporting the robustness of the mechanism formalized in Theorem~\ref{thm:delayed_peak}.

\subsection{Summary of Results}

The results support three central findings. First, behavioral recovery lags technical recovery due to bounded memory and broadcast persistence. Second, aggregate withdrawal pressure peaks after the outage nadir, often during recovery. Third, payment substitution reduces peak panic but does not guarantee reduced cumulative liquidity outflows unless paired with mechanisms that actively reduce rumor and scar.

\section{Discussion}
\label{sec:discussion}

This section interprets the results presented in Section~\ref{sec:results} and discusses their implications for payment system design, incident response, and financial stability analysis. We focus on mechanisms rather than numerical magnitudes, as the objective of the model is explanatory rather than predictive.

\subsection{Separation Between Technical and Behavioral Recovery}

A central finding of this work is the systematic separation between technical recovery and behavioral recovery. Even when payment success probabilities return to pre-incident levels, customer avoidance and withdrawal pressure remain elevated due to bounded memory, threshold effects, and persistent merchant broadcasts.

This separation implies that conventional operational metrics such as uptime, authorization success rates, or mean time to recovery are insufficient indicators of systemic risk. Behavioral variables evolve on a slower timescale and can dominate risk dynamics during the recovery phase. From a systems perspective, the incident does not end when technical metrics normalize; it ends when behavioral states stabilize.

\subsection{Why Peak Risk Occurs After ``Status Green''}

The delayed peak of withdrawal pressure identified in Theorem~\ref{thm:delayed_peak} and confirmed empirically challenges a common assumption in incident management: that the highest risk coincides with the worst technical performance.

In the model, withdrawals require sustained exposure to adverse signals and the crossing of multiple thresholds. These conditions are rarely met at the onset of an outage. Instead, they accumulate over time and often reach critical levels during recovery, when customers are still exposed to degraded merchant messaging and social confirmation while remaining operationally capable of moving funds.

This mechanism provides a formal explanation for why liquidity stress can intensify precisely when systems are declared recovered.

\subsection{Merchant Broadcasts as a Structural Amplifier}

Merchant behavior plays a distinct and structurally important role in the model. Broadcast persistence acts as a long-memory external field that sustains rumor formation independently of both social contagion and infrastructure reliability.

Operationally, this suggests that merchant guidance and signage are not merely secondary communication artifacts but first-order drivers of customer perception. In environments where merchant broadcasts are slow to normalize, technical investments in redundancy and failover may yield diminishing returns unless paired with coordinated merchant communication strategies.

\subsection{Interpreting the Role of Payment Substitution}

Payment substitution, modeled here as an instant transfer alternative to card payments, exhibits a dual effect. On the one hand, substitution consistently reduces peak avoidance and dampens panic by providing successful transaction experiences during periods of card degradation. On the other hand, substitution does not reliably reduce cumulative withdrawals under realistic broadcast persistence.

This apparent paradox arises because substitution preserves operational engagement while behavioral risk remains elevated. Customers who would otherwise disengage remain active and capable of executing withdrawals once eligibility thresholds are crossed. Substitution therefore mitigates acute panic but does not, by itself, eliminate latent run risk.

From a design perspective, substitution should be viewed as a panic-dampening mechanism rather than a complete risk control.

\subsection{Implications for Incident Response and Communication}

The model highlights the importance of post-recovery communication. Because rumor and scar decay slowly, silence or ambiguous messaging during recovery can be misinterpreted as concealment, reinforcing perceived systemic risk.

Effective incident response should therefore extend beyond technical remediation to include explicit reassurance, transparent postmortems, and coordinated merchant messaging. Such interventions can be interpreted within the model as mechanisms that accelerate the decay of rumor and scar, reducing the duration and magnitude of behavioral risk.

\subsection{Regulatory and Research Implications}

For regulators and supervisors, the results suggest that operational resilience frameworks should explicitly consider behavioral lag and information propagation. Stress testing that focuses solely on transaction volumes or settlement delays may underestimate tail risk if behavioral amplification channels are ignored.

From a research perspective, the model provides a bridge between bank run theory and operational risk analysis. It demonstrates how micro-level payment experiences can generate macro-level liquidity stress without requiring balance sheet deterioration or strategic coordination.

\subsection{Scope and Interpretation}

The findings should be interpreted as qualitative rather than quantitative predictions. The model does not attempt to replicate any specific institution or incident. Instead, it isolates mechanisms that are likely to operate across a wide range of digital payment environments.

Despite its abstractions, the consistency between theoretical results and simulation outcomes suggests that delayed behavioral risk is a robust feature of socio-technical payment systems rather than an artifact of specific parameter choices.

\section{Limitations and Future Work}

While the proposed model captures key behavioral and informational mechanisms linking payment outages to trust erosion and liquidity stress, several limitations should be acknowledged. These limitations also suggest directions for future research.

\subsection{Absence of Balance Sheet and Interbank Dynamics}

The model does not represent bank balance sheets, interbank exposures, or liquidity management actions. Withdrawals are treated as direct outflows without feedback from funding constraints, regulatory buffers, or central bank interventions. As a result, the model is not intended to predict solvency outcomes or systemic cascades in the interbank network.

Future work could integrate simplified balance sheet dynamics or couple the current behavioral model with existing interbank contagion frameworks to study feedback between depositor behavior and institutional liquidity management.

\subsection{Simplified Customer Decision-Making}

Customer agents are boundedly rational and follow threshold-based rules. While this structure captures nonlinear responses and hysteresis, it abstracts from strategic reasoning, learning, and heterogeneity in information processing beyond fixed parameter distributions.

Extensions could explore adaptive thresholds, learning-based updates, or alternative behavioral models informed by empirical data on payment usage and depositor responses during incidents.

\subsection{Stylized Representation of Merchant Behavior}

Merchant agents are modeled with simplified operational assessment and broadcast rules. Real-world merchant behavior is influenced by contractual obligations, acquirer guidance, and heterogeneous risk tolerance, which are not explicitly represented.

Future research could incorporate merchant–acquirer interactions, contractual incentives, or differentiated merchant types to better capture the diversity of acceptance behavior observed in practice.

\subsection{Exogenous Payment Infrastructure}

The payment infrastructure is treated as an exogenous stochastic process, independent of agent behavior. In reality, operational performance may degrade or recover endogenously due to load, retries, or risk controls triggered by customer actions.

Coupling infrastructure dynamics to agent behavior would allow the study of feedback loops between behavioral responses and technical performance, including overload-induced failures and recovery delays.

\subsection{Data and Calibration}

The model is not calibrated to any specific institution or historical incident. Parameters are chosen to reflect plausible ranges rather than empirical estimates. As such, numerical results should be interpreted qualitatively.

An important direction for future work is calibration and validation using incident logs, transaction-level data, or controlled experiments. Such data could be used to estimate memory decay rates, broadcast persistence, and substitution usage patterns.

\subsection{Additional Channels and Policy Interventions}

The current model focuses on card payments and a single substitution channel. Other channels, such as cash withdrawals, mobile wallets, or peer-to-peer payments, are not explicitly modeled. Similarly, policy interventions such as withdrawal limits, guarantees, or public communication strategies are not endogenized.

Future extensions could incorporate multiple channels and explicit policy actions to evaluate their relative effectiveness in mitigating behavioral risk and liquidity stress.

\section{Conclusion}

This paper investigates how payment outages and recoveries can generate delayed behavioral risk and liquidity stress through trust erosion, merchant signaling, and social contagion. Using a multi-agent simulation on a small-world network, we link micro-level payment experiences to macro-level withdrawal dynamics without relying on balance sheet deterioration or strategic coordination.

We show that bounded memory, threshold-based behavior, and persistent merchant broadcasts together imply a systematic separation between technical recovery and behavioral recovery. Under mild and realistic assumptions, aggregate withdrawal pressure can peak after the outage nadir, including during periods when payment systems have technically recovered. This delayed-peak mechanism is formalized analytically and validated empirically across a range of simulated scenarios.

Simulation results further demonstrate that payment substitution reduces peak panic but does not reliably eliminate cumulative liquidity outflows under realistic conditions. Substitution preserves operational continuity but does not, by itself, neutralize accumulated scar and rumor. These findings suggest that resilience mechanisms should be evaluated not only by their ability to restore functionality, but also by their impact on behavioral memory and perception.

The framework developed in this paper contributes to the literature on bank runs, operational resilience, and agent-based financial modeling by explicitly integrating payment reliability, merchant behavior, and social influence. By doing so, it highlights behavioral amplification channels that are typically outside the scope of traditional operational or liquidity stress testing.

From a practical perspective, the results underscore the importance of post-recovery communication, coordinated merchant messaging, and behavioral metrics in incident response. Declaring systems recovered does not imply that systemic risk has subsided. Managing the tail of behavioral risk is as critical as restoring technical performance.

Future work can extend this framework by incorporating balance sheet dynamics, endogenous infrastructure performance, and empirical calibration. Despite its abstractions, the model provides a tractable and extensible foundation for studying the behavioral dimensions of payment system resilience in increasingly digital financial environments.

\bibliographystyle{plain}
\bibliography{references}

\appendix
\section{Appendix}
\label{sec:appendix}

\subsection{Simulation Pseudocode}
\label{subsec:pseudocode}

Algorithm~\ref{alg:mas} summarizes the multi-agent simulation used throughout the paper.

\begin{algorithm}[t]
\caption{Multi-Agent Simulation of Payment Outage and Recovery}
\label{alg:mas}
\begin{algorithmic}[1]
\State Initialize customer agents with $(T_i(0), C_i(0), R_i(0), S_i(0)=\textnormal{OK}, B_i(0))$
\State Initialize merchant agents with $O_m(0)=B_m(0)=\textnormal{ACCEPTING}$
\State Generate Watts--Strogatz social network $G=(\mathcal{I},E)$
\For{$t = 1$ to $T$}
    \State Set infrastructure outcome probabilities $(p_s(t),p_f(t),p_u(t))$
    \For{each customer $i$}
        \State Draw payment attempt $A_i(t)$
        \If{$A_i(t)=1$}
            \State Select merchant $m \in \mathcal{M}_i$
            \State Draw outcome $Y_i(t)$
            \State Record experience signal $E_i(t)$
        \EndIf
    \EndFor
    \For{each merchant $m$}
        \State Update operational state $O_m(t)$
        \State Update broadcast state $B_m(t)$
    \EndFor
    \For{each customer $i$}
        \State Update $C_i(t), T_i(t), R_i(t), S_i(t)$
        \If{withdrawal eligibility satisfied}
            \State Draw withdrawal decision $W_i(t)$
            \If{$W_i(t)=1$}
                \State Reduce balance $B_i(t)$
            \EndIf
        \EndIf
    \EndFor
    \State Record system metrics
\EndFor
\end{algorithmic}
\end{algorithm}

\subsection{Model Parameters}
\label{subsec:parameters}

Table~\ref{tab:parameters} summarizes the main model parameters. All parameters are fixed within a simulation run but may vary across experiments.

\begin{table}[h]
\centering
\caption{Key model parameters and ranges}
\label{tab:parameters}
\begin{tabular}{lll}
\toprule
\textbf{Symbol} & \textbf{Description} & \textbf{Typical Range} \\
\midrule
$N$ & Number of customers & $10^3$--$10^4$ \\
$M$ & Number of merchants & $10^2$--$10^3$ \\
$k$ & Mean social degree & $6$--$12$ \\
$\beta$ & Network rewiring probability & $0.05$--$0.2$ \\
$\lambda_i$ & Baseline payment attempt rate & $[0.1,0.4]$ \\
$\rho_C$ & Scar memory persistence & $0.9$--$0.99$ \\
$\rho_T$ & Trust persistence & $0.8$--$0.95$ \\
$\rho_R$ & Rumor persistence & $0.9$--$0.99$ \\
$\gamma_C$ & Scar increment & $0.05$--$0.2$ \\
$\alpha_f$ & Failure severity weight & $0.5$--$1.0$ \\
$\alpha_u$ & Unknown severity weight & $0.8$--$1.2$ \\
$\theta_C$ & Scar threshold (withdrawal) & $0.4$--$0.7$ \\
$\theta_R$ & Rumor threshold (withdrawal) & $0.4$--$0.7$ \\
$\theta_i^{(1,2)}$ & Trust thresholds & heterogeneous \\
$\omega_i$ & Withdrawal fraction & $0.05$--$0.3$ \\
$\tau_m$ & Broadcast persistence time & $5$--$20$ steps \\
$D(t)$ & Demand multiplier & $1$--$2$ \\
\bottomrule
\end{tabular}
\end{table}

\subsection{Notes on Reproducibility}

All simulations are executed with explicit random seeds. Paired experiments (for example, with and without substitution) use identical seeds to isolate causal effects. Parameter values are selected to emphasize qualitative dynamics rather than calibration to a specific institution. The computational complexity per time step is linear in the number of agents and merchants,
O(N + M), as all updates are local and no global optimization or fixed-point computation is required.

\end{document}